# Orbital Domain Dynamics in a Doped Manganite

J. J. Turner[1,2*], K. J. Thomas[3], J. P. Hill[3], M. Pfeifer[1,2+], K. Chesnel[2++], Y. Tomioka[4], Y. Tokura[4,5], and S. D. Kevan[1]

[1]Dept. of Physics, University of Oregon, Eugene, Oregon 97403, USA; [2]Advance Light Source, Lawrence Berkeley National Laboratory, Berkeley, California 94720, USA; [3]Condensed Matter Physics Dept. and Material Science Dept., Brookhaven National Laboratory, Upton, New York 11973, USA; [4]Correlated Electron Research Center, National Institute of Advanced Industrial Science and Technology, Tsukuba 305-0033, Japan; [5]Department of Applied Physics, University of Tokyo, Tokyo 113-8656, Japan.

The coupling of multiple degrees of freedom – charge, spin, and lattice – in manganites has mostly been considered at the microscopic level. However, on larger length scales, these correlations may be affected by strain and disorder, which can lead to short range order in these phases and affect the coupling between them. To better understand these effects, we explore the dynamics of orbitally ordered domains in a half-doped manganite near the orbital ordering phase transition. Our results suggest that the domains are largely static, and exhibit only slow fluctuations near domain boundaries.

PACS numbers**:** 75.47.Lx, 71.27.+a, 64.60.Cn, 61.10.-i

In transition metal oxides, such as cuprates, nickelates and cobaltates, spin and charge order are intimately coupled. Neutron scattering studies that can compare spin and charge order as correlations grow with decreasing temperature provide important input for the theoretical models of these materials. For example, a comparison of spin and charge correlations ('stripe phases') in the copper-oxide plane of a doped cuprate [1], suggest that the correlations are coupled, while in similarly structured cobaltates the spin and charge correlations develop independently, forming a glass-like state [2].

In the half-doped manganites, three ordered phases – charge, orbital and magnetic - are coupled. In a useful, but oversimplified picture of these materials [3], there is one excess hole per two Mn sites. Below $T_{co}$ (or $T_{oo}$), these holes order into a checkerboard configuration [3] that creates a sub-lattice of $Mn^{3+}$ sites each with one electron in the degenerate $3d$ $e_g$ level (Fig. 1a). A cooperative Jahn-Teller (JT) distortion of oxygen anions surrounding the $Mn^{3+}$ sites lifts this degeneracy and sets the preference of the local orientation of the $e_g$ orbitals. Experimentally, charge and orbital order occur at the same temperature [4] ($T_{co} = T_{oo}$), while magnetic order occurs at or below [5,6] the charge/orbital-order transition.

In the simplest – and longest held – picture [3] of half-doped manganites [7], orbital and charge order [8,9,10] are seen as precursors to the magnetic ground state [3]. Experiments performed on a range of cubic manganites suggest that orbital order (OO) is short-ranged compared to magnetic and charge order [11,12], yet the origins of this short range order remain unclear. Strain fields [13], doping heterogeneity [14], and intrinsic structural defects [15] could all play a critical role in determining the short correlation length and orbital ordering dynamics near the phase transition. To address how dynamics might affect the correlations, we use a coherent resonant x-ray scattering technique that is sensitive to changes in the orbitally ordered domain state in the half-doped manganite $Pr_{0.5}Ca_{0.5}MnO_3$ (PCMO). We find that in the vicinity of the OO phase transition where the average domain size is decreasing with increasing temperature, the domains are largely static on the time scale of minutes accompanied by small-amplitude spatio-temporal fluctuations. We suggest that disorder may determine the short-range order as well as the *magnitude* of the domain-wall fluctuations. Our results indicate that studies in the time-domain are essential for a more complete portrait of the complex mesoscale physics of strongly-correlated electron systems.

While most neutron and x-ray experiments measure the atomic shifts associated with the JT-distortion, it is the OO electronic wave functions that determine the sign of the magnetic interaction between neighboring Mn sites [3,8]. To increase the sensitivity and selectivity of an x-ray diffraction measurement to the ordering of electronic orbitals, the x-ray energy can be tuned to near the Mn $L_{III}$ absorption edge (~ 650 eV), where the photon causes virtual electronic excitations between the Mn $2p$ core and $3d$ valence electronic levels [16]. Here the Mn form factor depends sensitively on the electronic structure of the $3d$ levels giving rise to scattering contrast between Mn sites with distinct orbital occupancies and orientations. The resonant enhancement permits the study of the correlations as they develop near the OO transition temperature [12,17] and, as we report here, enables the measurement of temporal fluctuations of the orbital domain boundaries.

Recent resonant x-ray scattering measurements challenge the conventional picture of OO in half-doped manganites. In particular, it was found that the average correlation length of the orbital domains is ~ 300 Å, about half the size of a typical magnetic domain [11] (~ 700 Å) and much less than the correlation length associated with charge order [4,18] (> 2000 Å). It remains unclear how to interpret these varying levels of mesoscale order between degrees of freedom that, in the simple model discussed above, are considered to be closely coupled.

In this Letter, we address these questions by looking more closely at the orbitally ordered domains themselves. We used longitudinally and transversely coherent x-rays tuned to the Mn L-edge to measure an orbital order diffraction peak in a $Pr_{0.5}Ca_{0.5}MnO_3$ (PCMO) single crystal. With coherent illumination, constructive and deconstructive interference occurs between x-rays diffracted from different orbitally ordered domains, and the resultant scattering exhibits a characteristic speckle pattern [19] (Fig. 1b). We exploit the fact that changes in the real space domain configuration are reflected by changes in this speckle pattern to monitor orbital domain dynamics as a function of temperature and time. The main analysis therefore involves correlating speckle patterns from the same position in reciprocal space, but separated in time. Although similar measurements were attempted at the Mn $K$-edge (6.55keV), the signal rates were too low to yield conclusive results [20]. Using soft x-rays, much higher counting rates are expected in dynamics experiments [21] and a direct measurement of orbital domains [22] is allowed.

The PCMO sample exhibits two transitions, a simultaneous orbital- and charge-ordering transition near T ~ 235 K and an antiferromagnetic transition at T ~ 170 K [5,6]. Using a orthorhombic lattice unit cell, the reflection takes place at (0 ½ 0), giving a 2θ scattering angle of ~ 124° at the Mn $L$-edge. This experiment was performed on beamline 12.0.2.2 of the ALS at the Lawrence Berkeley National Laboratory. The system allows us to control the temperature of the sample to 0.01K. The detector used was a 2048 x 2048 pixel charge-coupled device (CCD) with 12.5 square micron pixels.

The partially coherent x-ray source is prepared by passing the x-ray beam through a ~ 10 micron sized pinhole which is ~ 3 mm in front of the sample (Fig. 1). The transverse coherence length of the beam at the sample is ~ 5 microns which means we coherently illuminate ~ $10^5$ orbital domains. For a fixed temperature, we recorded a sequence of speckle patterns at the (0 ½ 0) diffraction peak separated by equal increments of time. The images shown in Fig. 2(a) and 2(c) were collected at T=200K and T=232K, well below and near $T_{oo}$, respectively. Along with the expected intensity decrease, the envelope of the peak is seen to broaden at the higher temperature, indicating a reduced average OO domain size. Large-scale domain wall motion should lead to changes in the speckle pattern which would be apparent in the time scans, shown for the two temperatures in Fig. 2(b) and 2(d). These plot the evolution of the speckle intensity as a function of time for a vertical cut through the middle of the Bragg peak. Surprisingly, both time scans show straight horizontal lines, indicating that the domains are largely static on the time scale of the data set.

We fit the orbital peak to a Lorentzian function to obtain a metric of the intensity and correlation length systematically as a function of temperature (Fig. 3a). The orbital domain size obtained by these fits is consistent with recent measurements on PCMO samples of different dopings [12] as well as other manganites [17]. The two plots clearly delineate domain melting as $T_{oo}$ is approached from below. The broadening of the envelope is consistent with a decrease in the average size of the orbital domains, while the decrease in integrated intensity suggests the volume of orbital order has also decreased. In addition, we observed the onset of a constant, isotropic background that appears as the correlation length starts to change and saturates above the transition. We attribute this background to correlated polarons which have been observed previously [23] and which produce a diffuse scattering signal in our experiment. This background persists well above 300 K, consistent with earlier studies [23].

To carry out a more detailed search for dynamics, we systematically raised the sample temperature in one degree increments through the transition, waiting for one hour for the system to stabilize at each point. At each temperature, we acquired a series of 100 speckle patterns, each image corresponding to a 10s exposure, for a total of 20 minute sets. For each temperature T, we then calculated momentum-resolved cross-correlation coefficients $<\rho(\mathbf{q},\tau,T)>$ to map fluctuations of orbital domains. For proper normalization, $\rho = 1$ for two perfectly correlated speckle patterns, and $\rho = 0$ for completely uncorrelated patterns. Computations performed with a similar normalization were calculated by Pierce *et. al.* [24]. Here however, to get $\rho(\tau)$, the coefficient *change* with $\tau$ must be calculated. For $\tau$-dependence, coefficients are averaged over all cross-correlations between pairs of images separated by the same delay time $\tau$. The **q**-component is computed by using the coordinate system of the CCD camera and transforming each data point to its precise $\mathbf{Q} = \mathbf{k}_f - \mathbf{k}_i$ in momentum space, where the **k**-vectors are the initial and final wavevectors of the photons (Fig. 1). The difference **q,** of this **Q**-value and the reciprocal lattice vector **G** forms annuli of constant magnitude centered on the orbital Bragg peak. Correlating the speckle pattern within those regions of constant **q**, $<\rho(\tau,\mathbf{q})>$ is

computed for each region and can be incorporated into a 2D map as a function of both **q** and $\tau$:

$$\rho(\mathbf{q},\tau)=\sum_{t=1}^{n} \frac{1}{N(t,\tau)^{1/2}} A(\mathbf{q},t) \wedge B(\mathbf{q},t+\tau)$$

where n is the number of images, $\wedge$ represents a cross-correlation operation, and N(t,$\tau$) is a normalization constant that is a function of both real time t, and image separation time, and is equal to:

$$N(t,\tau)=[n-\tau]^2 * [A(\mathbf{q},t) \wedge A(\mathbf{q},t)] * [B(q,t+\tau) \wedge B(q,t+\tau)]$$

Each map consists of all permutations of pairs of images within a 100-image set, and multiple image sets per temperature are averaged to finally get <$\rho$(**q**,$\tau$,T)>.

A sampling of the correlation map results is presented in Fig. 4. For T < 231K, <$\rho$(**q**,$\tau$,T)> is very close to one at all delay times and wave vectors measured, confirming that the low-T speckle patterns are in fact static and, not incidentally, that the experimental set-up is stable enough for the detailed measurements outlined here. The correlation maps for T = 232 K and T = 233 K show small, but measurable deviations from unity. The **q**-dependence shows smaller length scales fluctuating faster, as is expected. Two key points arise from this analysis: first, that only a small fraction of the domains structure is fluctuating; and second, that this fraction comprises physics with slow, measurable dynamics near the transition.

The first can be seen as even at the longest times measured, <$\rho$($\tau$,**q**,T) > does not decay by more than 10%. The system behaves as though it has frozen-in disorder that prevents sizeable fluctuations on the timescale of the experiment. Although x-rays at these energies only penetrate 500-1000 Å into the sample, we believe it is unlikely that the pinning is due to defects at the surface. In particular, we note that the size of the domains observed here and the temperature dependence of the order parameter are both consistent with the results of neutron and hard x-ray measurements, which probe the bulk of the material [4,18,20,25]. In addition, short-range OO, with orbital domains of ~ 300-500 Å, appears to be a general result from *L*-edge scattering studies on a variety of manganite materials prepared in different laboratories [11,12,17]. These arguments suggest that a more general mechanism is responsible for pinning the of domain walls. For instance, domain walls are expected to be pinned at points where the cost in elastic energy for having neighboring OO wavefunctions out of phase is lowest. One candidate for such a site is the Ca dopant ions, though the fact that the domains are typically 100 lattice constants in size seems inconsistent with the high calcium concentration in our samples. Statistical calcium concentration variations might well produce enough local strain to pin domain walls and, while also occurring on the correct length scale, to rationalize the observed OO domain size.

Regarding the second point, the small-amplitude fluctuating component that emerges only near the phase transition (Fig. 4) is puzzling. These dynamics are related to limited domain structure motion and exhibit fluctuations that occur on the order of hundreds of seconds. The long time scale of this measurement is therefore slower than any measured mechanism that we know of at this time. These results could provide a clue to understanding orbital ordering in the manganites through slow dynamics on large length scales, congruent to orbital correlation lengths.

An interpretation that is consistent with our results suggests a picture in which at a given temperature, the bulk of a domain is static. The domain walls are pinned except for temperatures very close to the transition. Past some dynamic threshold near the transition, the domain walls are able to execute small-amplitude motion. This behavior contrasts, for example, with the domain wall dynamics associated with charge density waves in the simple antiferromagnet chromium [26]. In chromium, the domains are a few microns in size and fluctuate on a length scale comparable to the domain size itself, resulting in speckle patterns that become completely uncorrelated at long times [26]. Thus, it appears that in contrast with these systems, the manganites possess a source of quenched disorder which couples to the OO order parameter. This coupling pins the domain walls, and may prove the manganites to be analogous to recent observations in the displacive, $SrTiO_3$ antiferrodistortive transition [27]. However, the apparent degree of freedom in orbital domain wall motion is a completely novel observation, unlike either of these systems, and we have not found an analog.

In the manganites and other complex oxides, the interactions between order parameters are complex and involve a range of length scales. The results here show that a great deal of information can be collected by probing these systems in the time-domain. In particular, we have shown that the short-range orbital correlations are largely static with only small fluctuations as the transition is approached. This highlights the importance of quenched disorder in this system for determining its ground state properties. If the domain wall dynamic measurements reported here can be applied to other complex oxides, dynamic stripe correlations in the high-temperature superconductors [28] for instance, then many forms of time-dependent heterogeneity can be readily probed.

This work was supported in part by the National Science Foundation under grant DMR-0506241. The Advanced Light Source is operated under Contract DE-AC03-76SF00098 at Lawrence Berkeley National Laboratory and is supported by the US Department of Energy.

*Author to whom correspondence should be addressed: J. J. Turner (jjturner@lbl.gov).

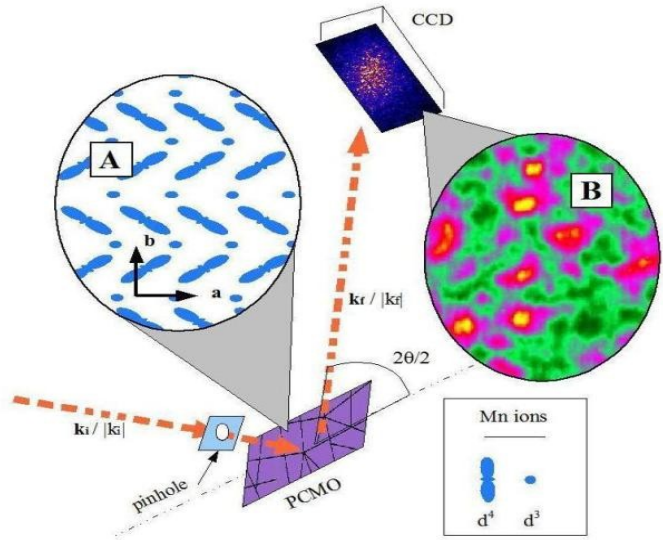

**Fig. 1. Coherent scattering experimental set-up** The scattering geometry is set to measure the orbital Bragg condition at **Q** = (0, 1/2, 0) in $Pr_{0.5}Ca_{0.5}MnO_3$. The cartoon (inset A) depicts the CE-type orbital-ordering of the $e_g$ electron of Mn for a single domain, which originates in the a-b plane. The key given at the bottom of the figure shows the symbols used for the $d^4$ and $d^3$ ions in the inset. Orbital ordering takes place only between ions with the $d^4$ electronic structure. Coherent interference between scattered waves from the domain structure of the sample results in a 'speckle' pattern (inset B). Shown is a close-up view of the image taken on a CCD camera from a region of the image, demonstrating the well-developed speckle and exceptional contrast at about ~ 80%.

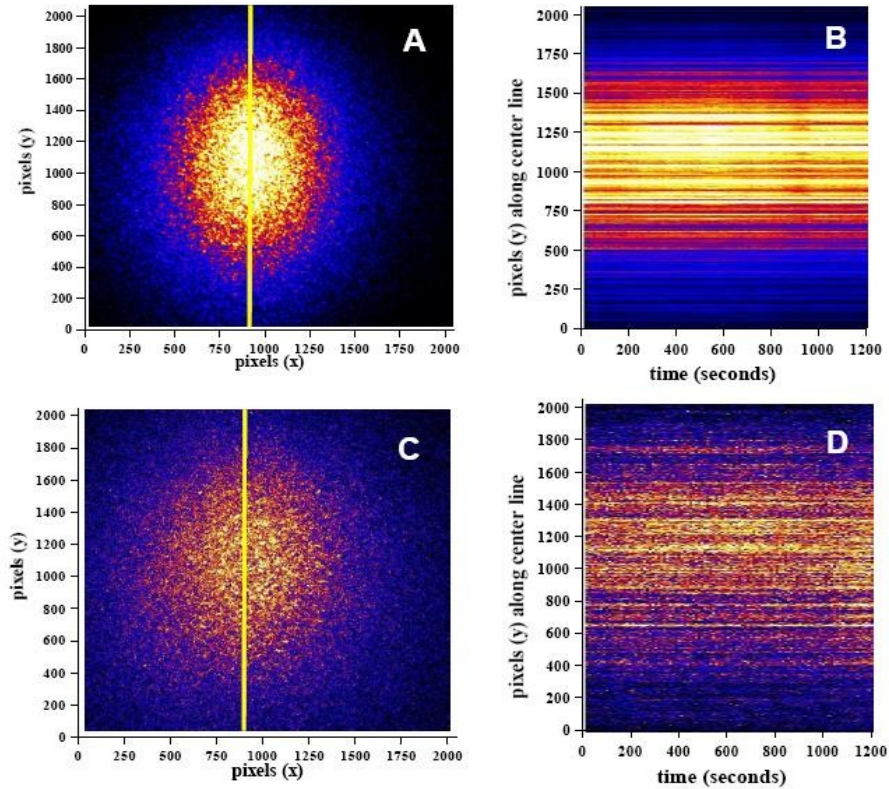

**Fig. 2. Time evolution of speckle** (A) Speckle pattern taken at the orbital Bragg peak deep within the orbital ordering phase at T = 205 K. The exposure time for this image was 10s. (B) Time evolution of the speckle pattern in (A) for a vertical slice (yellow line) through the center of the image. This figure represents the evolution of a 1d scan through the image like that shown in (A) of a 100-image set, taken over a 20 minute period. The speckle structure is static as evidenced by the straight horizontal lines. (C) and (D) The same as (A) and (B) for T = 232 K, near the OO phase transition.

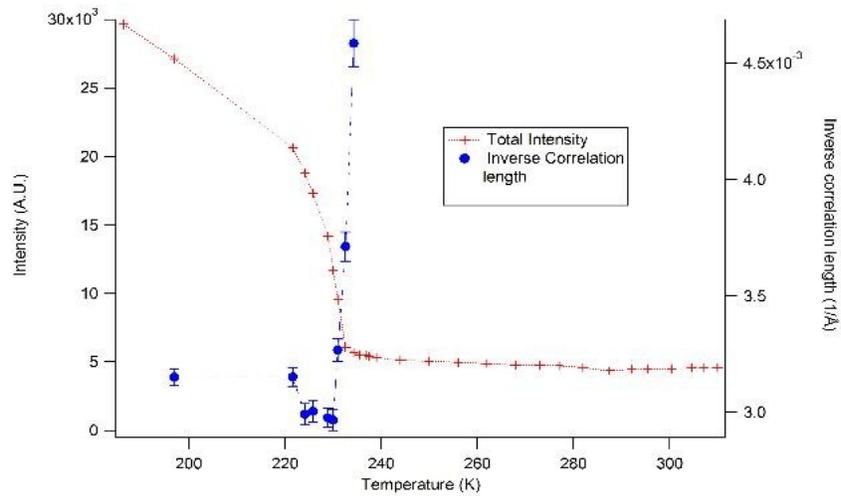

**Fig. 3. Near-transition behavior** The integrated intensity for the orbital ordering mechanism (red) and average domain size (blue) as a function of temperature. The domain volume starts to decrease as the temperature approaches the transition from below and the width of the orbital peak, or inverse correlation length, increases. Well below the transition, the orbital domain size is on the order of 300Å.

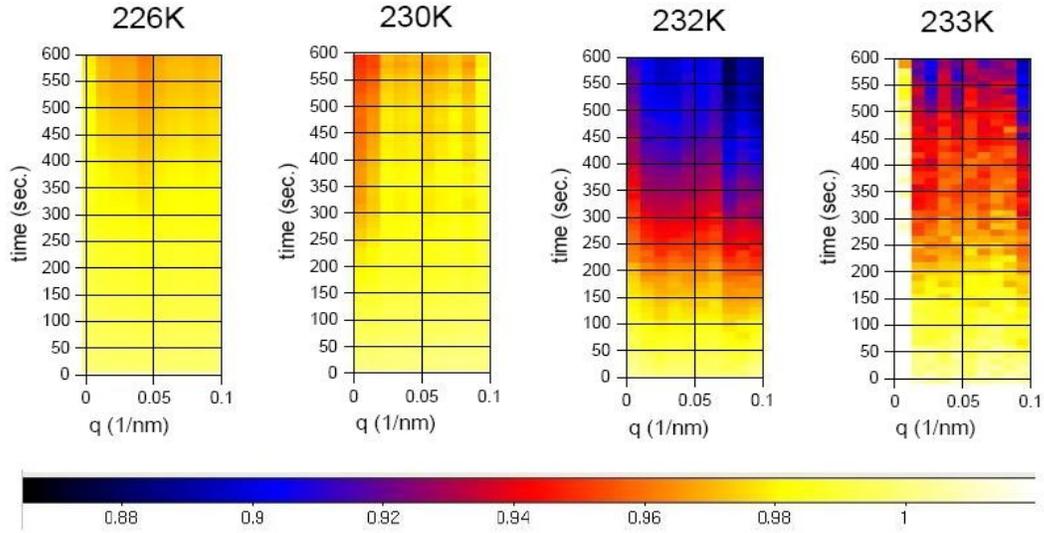

**Fig. 4. Cross-correlation coefficient maps** Each frame represents the average of correlations for three separate maps for a given temperature. Time is calculated by averaging over all permutations of pairs separated by delay time $\tau$. The quantity **q** was calculated by computing all three components of the wave vector transfer **Q** – **G** and taking the magnitude, where the wavevector magnitude $|\mathbf{k}|$ is $2\pi/\lambda$, **Q** is given by $\mathbf{k_r} - \mathbf{k_i}$, and **G** is the orbital ordering reciprocal lattice vector. The **q**-resolution is 0.01 nm$^{-1}$ per pixel. Even for the **q** and $\tau$ dependence shown above, the absolute changes are small.